\documentclass[floatfix,aps,prd,eqsecnum,showpacs,amsmath,amssymb,nofootinbib,preprintnumbers]{revtex4}

\usepackage{graphicx}
\usepackage{color}

\usepackage{hyperref}
\hypersetup{colorlinks=true,linkcolor=blue,pdfstartview=XYZ}

\def\laq{~\raise 0.4ex\hbox{$<$}\kern -0.8em\lower 0.62ex\hbox{$\sim$}~}
\def\gaq{~\raise 0.4ex\hbox{$>$}\kern -0.7em\lower 0.62ex\hbox{$\sim$}~}

\def\beq{\begin{equation}}
\def\eeq{\end{equation}}
\def\bea{\begin{eqnarray}}
\def\eea{\end{eqnarray}}

\def \pa {\partial}

\def \ti {\widetilde}

\def \ga {\gamma}

    \def\be{\begin{equation}}
    \def\ee{\end{equation}}
    \def\ba{\begin{eqnarray}}
    \def\ea{\end{eqnarray}}

\newcommand{\eq}{\begin{equation}}
\newcommand{\eqx}{\end{equation}}
\newcommand{\eqn}{\begin{eqnarray}}
\newcommand{\eqnx}{\end{eqnarray}}

\newcommand{\Ups}{\Upsilon}

\newcommand{\Hcal}{\mathcal H}

\usepackage{pifont}

\begin{document}

\title{The luminosity distance-redshift relation up to second order in the Poisson gauge with anisotropic stress}

\author{Giovanni Marozzi$^{1}$}

\affiliation{
 $^{1}$Universit\'e de Gen\`eve, D\'epartement de Physique Th\'eorique and CAP,
24 quai Ernest-Ansermet, CH-1211 Gen\`eve 4, Switzerland
}



\begin{abstract}
We present the generalization of previously 
published results, about the perturbed redshift and the luminosity-redshift 
relation up to second order in perturbation theory, for the case of the Poisson gauge and in the presence of anisotropic 
stress. The results are therefore valid for general dark energy models and (most) modified gravity models.
We use an innovative approach based on the recently proposed "geodesic light-cone" gauge.
We then compare our finding with other results, which recently appeared in the literature, for the particular 
case of vanishing anisotropic stress.
Arriving at a common accepted expression for the non-linear and relativistic corrections  to the redshift and distance-redshift relation is of fundamental importance in view of future cosmological 
surveys. Thanks to these surveys the Universe will be further probed with high precision and at very different scales, where non-linear and relativistic effects can play a key role. 
\end{abstract}

\vspace {1cm}~


\maketitle

\section{Introduction}
\label{Sec1}
\setcounter{equation}{0}

In the near future cosmology will enter a new era in which the use of Newtonian gravity  
will no longer be sufficient in studying large scale structure (LSS). In fact, the next generation of LSS surveys will probe the Universe with high precision and at very different scales, where 
non-linear and  relativistic effects can play a key role.
Therefore, it is of fundamental importance to have a reliable description of the observables which describe the physical information carried by light-like signals traveling along our past light-cone, at least up to second order in perturbation theory. 
Among these observables the redshift $z$ and the luminosity distance $d_L$ occupy important positions. In fact,
following the pioneering work of \cite{Sasaki}, $d_L$ has been computed to first order in the longitudinal gauge (for a CDM model in \cite{Bonvin:2005ps}, CDM and $\Lambda$CDM in \cite{Pyne:2003bn}), and to second order in the synchronous gauge, but only for CDM, in \cite{Barausse:2005nf}.

Here we generalize the results presented in \cite{BenDayan:2012wi,Fanizza:2013doa} (and used in \cite{BenDayan:2012ct,BenDayan:2013gc,Ben-Dayan:2014swa}), where the perturbed redshift and luminosity distance-redshift relation were obtained up to second order in the Poisson gauge and for a general dark energy model but with vanishing anisotropic stress,  to the case where  anisotropic stress is present.

The evaluation of LSS observables in the presence of anisotropic stress is one of the major issue to be addressed in view of the next generation of LSS surveys. In fact the anisotropic stress, which vanishes for the  $\Lambda$CDM model, 
frequently appears in other dark energy  and/or modified gravity models. The main point is that the presence of anisotropic stress can induce deviations from the standard observational predictions based on $\Lambda$CDM. Therefore, if we are able to isolate and measure the effect of the anisotropic stress in future LSS cosmological observations we will be able to conclude that the Universe is not described by a  $\Lambda$CDM model.

In \cite{BenDayan:2012wi,Fanizza:2013doa}  the perturbed redshift and luminosity-redshift relation were derived for the first time up to second order in the Poisson gauge, and for a general dark energy model, starting from the recently proposed "geodesic light-cone" (GLC) gauge \cite{Gasperini:2011us} and using an innovative approach. 
On the other hand, the final results of~\cite{BenDayan:2012wi,Fanizza:2013doa} are valid only for the case with vanishing anisotropic stress and are only partially written using a formalism which is simply related to the one already used at first order (see, for example, \cite{Bonvin:2005ps}).
Here we fill this gap.

For problems associated with the observation of light sources lying on the past light-cone of a given observer, the GLC gauge, an adapted system of coordinates, is extremely helpful. In this system several quantities simplify greatly~\cite{Gasperini:2011us} and the so-called Jacobi Map can be obtained exactly, non perturbatively, \cite{Fanizza:2013doa}, while keeping all the required degrees of freedom for applications to general geometries.
As a consequence, starting from the GLC gauge one can express light-cone observables in any gauge 
by computing a coordinate transformation that connects the GLC to the chosen gauge (see \cite{BenDayan:2012wi,Fanizza:2013doa} for details). This new procedure considerably simplifies
the task of writing LSS observables (like redshift and luminosity distance) to a given order in perturbation theory.
In practice, one can start from a given non-perturbative exact expression for the observable in question in GLC gauge, and go to its perturbative counterpart, e.g. in Poisson gauge, using a coordinate transformation valid at the desired order in the perturbative theory.

In the second part of the paper
we also attempt a comparison of the results of~\cite{BenDayan:2012wi,Fanizza:2013doa}, and of the ones here presented,
with other results, most notably~\cite{Umeh:2014ana} (see also \cite{Umeh:2012pn}), for the case of vanishing anisotropic stress.
As we shall see, even after translation the comparison is not straight-forward due to the length of the expressions and the possibility of transforming them by integrations by parts. The result is simply that the expressions derived here and 
in~\cite{BenDayan:2012wi,Fanizza:2013doa} do not agree with the ones derived in~\cite{Umeh:2014ana}. Further work will be needed to resolve the discrepancies. In order to encourage colleagues to look further into this comparison, we believe it is useful to include here this first attempt.

The paper is organized as follows. In Section 2 we recall the definition and special properties of  the geodesic light-cone gauge.
We also specify the Poisson gauge up to second order in perturbation theory for the case with anisotropic stress, and find the connection between the two gauges up to second order.
In Section 3, we first give the result for the redshift up to second order in perturbation theory in the Poisson gauge in terms of the observer's angular coordinates, and for a generic dark energy (modified gravity) model with anisotropic stress, using the standard formalism. We then move to the luminosity distance as function of the observed redshift and of the observer's angular coordinates, also up to second order in perturbation theory in the Poisson gauge and for a generic dark energy (modified gravity) model with anisotropic stress.
In Section 4 we consider the particular case of vanishing anisotropic stress and compare our results (and the ones of  
\cite{BenDayan:2012wi,Fanizza:2013doa}) with the results of \cite{Umeh:2014ana,Umeh:2012pn}.
In Section 5 we summarize our results and draw some conclusions.

\section{From the Geodesic Light-Cone to the Poisson gauge}
\label{Sec2}
\setcounter{equation}{0}

Following \cite{BenDayan:2012wi,Fanizza:2013doa} we give in this paper the expression for the redshift and the luminosity distance-redshift relation in a generic homogeneous FLRW Universe with perturbation, up to second order and in the Poisson gauge with anisotropic stress.
We start with the so-called geodesic light-cone (GLC) coordinate defined in \cite{Gasperini:2011us}.
GLC coordinates consist of a timelike  coordinate $\tau$ (which can always be  identified with the proper time of the synchronous gauge and, 
therefore, describes a geodesic observer static in this gauge \cite{BenDayan:2012pp}),  of a null coordinate $w$ and of two angular coordinates $\tilde{\theta}^a$ ($a=1,2$). 

The line-element of the GLC metric takes the form:  
\beq
\label{LCmetric}
ds^2 =\Upsilon^2 dw^2 - 2 \Upsilon  dw d\tau+\gamma_{ab}(d \tilde{\theta}^a-U^a dw)(d \tilde{\theta}^b-U^b dw) ~~,~~~~~ a, b = 1,2~~,
 \eeq
 and depends on six arbitrary functions ($\Upsilon$,  $U^a$ and $\gamma_{ab} = \gamma_{ba}$).
In matrix form:
\beq
\label{GLCmetric}
g_{\mu\nu} =
\left(
\begin{array}{ccc}
0 & -\Upsilon &  \vec{0} \\
-\Upsilon & \Upsilon^2 + U^2 & -U_b \\
\vec0^{\,T}  &-U_a^T  & \gamma_{ab} \\
\end{array}
\right)
~~~~~,~~~~~
g^{\mu\nu} =
\left(
\begin{array}{ccc}
-1 & -\Upsilon^{-1} & -U^b/\Upsilon \\
-\Upsilon^{-1} & 0 & \vec{0} \\
-(U^a)^T/ \Upsilon & \vec{0}^{\, T} & \gamma^{ab}
\end{array}
\right) ~,
\eeq
where $\gamma_{ab}$ and its inverse  $\gamma^{ab}$ lower and  raise  the two-dimensional indices~\footnote{However, in analogy with the 
synchronous gauge, also the GLC gauge has some residual gauge freedom \cite{Fanizza:2013doa}.}. 

The condition $w=$ constant defines a null hypersurface ($\pa_\mu w \pa^\mu w=0$), corresponding to the past light-cone of the given observer,
hereafter chosen to be the geodesic one.  
The vector $u_\mu = - \partial_{\mu} \tau$ is the 4-velocity of this geodesic observer, $\left( \pa^\nu \tau\right) \nabla_\nu \left( \pa_\mu \tau\right) = 0$. Let us also recall  that, in GLC gauge, the null geodesics connecting sources and observer are characterized simply by the tangent vector $k^{\mu} = - \omega g^{\mu \nu} \partial_{\nu} w = -  \omega g^{\mu w} = \omega  \Upsilon^{-1} \delta^{\mu}_{\tau}$ (where $ \omega$ is an arbitrary normalization constant), meaning that photons  travel at constant values of $w$ and $\tilde{\theta}^a$. 
This renders the calculation of  the redshift particularly simple in this gauge.

We now determine the redshift and the cosmological distances, as the luminosity and angular distance, exactly, non-perturbatively  in GLC gauge. Let us  denote by subscripts ``o'' and ``s'', respectively,  quantities evaluated at the observer and source space-time position, and let us consider a light ray emitted by a static geodesic source lying on the past light-cone of a static geodetic observer (defined by $w=w_o$) and on the spatial hypersurface $\tau= \tau_s$.
The light ray will be received by the static geodetic observer at $\tau=\tau_o>\tau_s$. 
The exact non-perturbative expression of  the redshift $z_s$ associated with this light ray is then simply given by \cite{Gasperini:2011us}
\be
\label{redshift}
(1+z_s) = \frac{(k^{\mu} u_{\mu})_s }{(k^{\mu} u_{\mu})_o}  = \frac{(\partial^{\mu}w \pa_\mu \tau)_s }{(\partial^{\mu}w \pa_\mu \tau)_o}  = {\Ups(w_o, \tau_o, \ti \theta^a)\over \Ups(w_o, \tau_s, \ti \theta^a)} ~~.
\ee

On the other hand, in \cite{Fanizza:2013doa} an exact expression for the so-called Jacobi Map~\cite{SEF} is derived in GLC gauge and the following non-perturbative solution for the luminosity (area) distance is obtained:
\beq
\label{lumdist}
d_L^2 =  (1+z_s)^4 d_A^2 = 
(1+z_s)^4 \frac{4 \sqrt{\gamma_s}}{\left[\det \left(u_{\tau}^{-1} \pa_\tau{\gamma}^{ab}\right) \gamma^{3/2}\right]_{o}} \,, 
\eeq
where $\gamma$ denotes the determinant of the 2-dimensional matrix $\gamma_{ab}$.
 
Let us now define the gauge in which we want express the redshift and the luminosity distance, given by Eqs.(\ref{redshift}) and (\ref{lumdist}).
Neglecting  vector and tensor contributions, the Poisson gauge (PG) metric~\cite{PG} (sometimes denoted at first order 'Newtonian gauge' or 'longitudinal gauge')
takes the  form
\bea
ds_{PG}^2 &=& a^2(\eta) \left[ -(1+ 2 \Phi) d\eta^2  + (1- 2 \Psi)\delta_{ij}  dx^i dx^j \right]  \nonumber \\
&=&  a^2(\eta) \left[ -(1+ 2 \Phi) d\eta^2  + (1- 2 \Psi)( dr^2 + r^2 d^2 \Omega) \right] 
\label{PGmetricstandard}
\eea
where the (generalized) Bardeen potentials $\Phi$ and $\Psi$ are defined, up to second order, as follows:
\be
\Phi \equiv \phi + \frac{1}{2} \phi^{(2)} ~~,~~~~~~~ \Psi \equiv \psi + \frac{1}{2} \psi^{(2)} ~~,
\ee
and we make no assumption on the anisotropic stress, so that $\Psi$ amd $\Phi$ can be different also at first order. 

In order to compute the redshift and the luminosity distance given in Eqs.(\ref{redshift}) and (\ref{lumdist}) in terms of standard PG variables we have to transform the GLC gauge quantities to quantities in PG. 
This generalize what done in \cite{BenDayan:2012wi,Fanizza:2013doa}, 
because we consider here the general case with anisotropic stress up to second order.
Starting from the following suitable boundary conditions:
$i)$ the transformation is non singular around $r=0$, and $ii)$  the two-dimensional spatial section $r=$ const are locally parametrized at the observer position by standard spherical coordinates $(\theta,\phi)$, 
the coordinate transformation to second order  
is then given by
\bea
\tau &=& \tau^{(0)}+\tau^{(1)}+\tau^{(2)} \nonumber \\
& & \text{with} \quad \tau^{(0)} = \left( \int_{\eta_{in}}^\eta d\eta' a(\eta') \right) \quad,\quad \tau^{(1)} = a(\eta) P(\eta, r, \theta^a)\,, \nonumber \\
& & \quad \quad \quad \tau^{(2)}= \int_{\eta_{in}}^\eta d\eta' \frac{a(\eta')}{2} \left[ \phi^{(2)} - \phi^2 + ( \partial_r P )^2 + \gamma_0^{ab} ~ \partial_a P ~ \partial_b P \right] (\eta', r, \theta^a)~,
\label{tau2order} \\
\!w \!&=&\! w^{(0)}+w^{(1)}+w^{(2)} \nonumber \\
\!& &\!\text{with} \quad w^{(0)}=\eta_+ \quad,\quad w^{(1)}=Q(\eta_+, \eta_-, \theta^a) \,,
\nonumber \\
& & \quad \quad \quad 
w^{(2)}= { \frac{1}{4} \int_{\eta_o}^{\eta_-} dx~ \left[ {\psi}^{(2)} + {\phi}^{(2)}+2(\psi^2-\phi^2) + 2 (\psi+\phi) \partial_+ Q + {\gamma}_0^{ab} ~ \partial_a Q ~ \partial_b Q \right] (\eta_+, x, \theta^a)}\,,
\label{w2order} \\
\!\tilde{\theta}^a \!&=&\! \tilde{\theta}^{a (0)}+\tilde{\theta}^{a (1)}+\tilde{\theta}^{a (2)} \nonumber \\
\!& &\!\text{with} \quad \tilde{\theta}^{a (0)}=\theta^a 
\quad,\quad
\tilde{\theta}^{a (1)}={\frac12 \int_{\eta_o}^{\eta_-} dx~ \left[ {\gamma}_0^{ab} \partial_b Q \right] (\eta_+,x,\theta^a)} 
\nonumber \\
& & \quad \quad \quad 
\tilde{\theta}^{a (2)}= { \int_{\eta_o}^{\eta_-} dx~ 
\left[ 
\frac{1}{2}
{\gamma}_0^{ac} \partial_c w^{(2)} 
+ {\psi} {\gamma}_0^{ac} \partial_c w^{(1)} 
+\frac{1}{2}{\gamma}_0^{dc} \partial_c w^{(1)} \partial_d   \tilde{\theta}^{a (1)}
+\frac{1}{2}(\psi+\phi)\partial_+\tilde{\theta}^{a (1)}
+(\phi-\psi) \partial_-\tilde{\theta}^{a (1)}
\right. } \nonumber \\
& & \left. \quad \quad\quad\quad\quad\quad\quad\quad\quad\quad\,\,\,\,
-\partial_+ w^{(1)} \partial_-\tilde{\theta}^{a (1)} \right](\eta_+,x,\theta^a) \,,
\label{thetatilde2orderShort}
\eea
where $(\ga_0^{ab}) = {\rm diag}(r^{-2},r^{-2} \sin^{-2}\theta)$, and $\eta_{in}$ represents an early enough time when the perturbations (or better their integrands) were negligible.
We have also introduced the zeroth-order light-cone variables $\eta_\pm= \eta \pm r$, 
with corresponding partial derivatives:
\beq
\pa_\eta = \pa_+ + \pa_- ~~~,~~~~~ \pa_r = \pa_+ - \pa_- ~~~,~~~~~\pa_\pm= {\pa \over \pa \eta_\pm}={1\over 2} \left( \pa_\eta \pm \pa_r \right) ~~,
\eeq
and defined
\be
P(\eta, r, \theta^a) = \int_{\eta_{in}}^\eta d\eta' \frac{a(\eta')}{a(\eta)} \phi(\eta',r,\theta^a)
\,\,\,\,,\,\,\,\, ~~~ Q(\eta_+, \eta_-, \theta^a) = \int_{\eta_o}^{\eta_-} dx~ \frac{1}{2}
\left(\psi+\phi\right)(\eta_+,x,\theta^a) ~.
\label{PQ}
\ee
With this we can then compute the non-trivial entries of the GLC metric of Eq. (\ref{GLCmetric}) in terms of the variable $(\eta,r,\theta^a)$:
\bea
\Upsilon^{-1} &=& \frac{1}{a(\eta)} \left[ 1 + \partial_+ Q - \partial_r P  +\frac{1}{2}(\psi-\phi)
+\partial_{\eta}w^{(2)} + \frac{1}{a}(\partial_\eta - \partial_r) \tau^{(2)} - \phi^{(2)} + 2\phi^2 
-\frac{1}{2}\phi(\phi+\psi)
\right. \nonumber \\
& & \left.
-\phi \partial_+ Q+\partial_r P  \left(\frac{1}{2}\phi-\frac{3}{2}\psi\right)- \partial_r P\partial_+ Q
- \gamma^{ab}_0 \partial_a P \partial_b Q\right] \,,
\label{Ups1}  
\\
U^a &=& \partial_{\eta}\tilde{\theta}^{a (1)}-\frac{1}{a}\gamma_0^{ab}\partial_b \tau^{(1)}+\partial_{\eta}\tilde{\theta}^{a (2)}-
\frac{1}{a}\gamma^{ab}_0\partial_b \tau^{(2)} - \frac{1}{a} \partial_r \tau^{(1)} \partial_r \tilde{\theta}^{a(1)}  \cr
& &  -\phi \partial_\eta \tilde{\theta}^{a (1)}-\frac{2}{a} \psi \gamma^{ab}_0 \partial_b \tau^{(1)} 
-\frac{1}{a}\gamma_0^{cd}\partial_c \tau^{(1)} \partial_d \tilde{\theta}^{a (1)} \nonumber \\
& &
+\left(\partial_+ Q - \partial_r P+\frac{1}{2}(\psi-\phi)\right) \left(-\partial_{\eta} \tilde{\theta}^{a (1)}+\frac{1}{a}\gamma^{ab}_0 \partial_b \tau^{(1)}\right)\,,
\label{Ua1}
\\ 
\gamma^{ab} &=& a^{-2}\left\{ \gamma_0^{ab} \left(1 +  2 \psi\right) +\left[\gamma_0^{a c} \partial_c \tilde{\theta}^{b (1)}+ (a\leftrightarrow b) \right]+ \gamma_0^{ab}
\left(\psi^{(2)} + 4 \psi^2 \right)-\partial_\eta \tilde{\theta}^{a (1)}\partial_\eta \tilde{\theta}^{b (1)}
\right. \nonumber \\
& & \left.
+\partial_r \tilde{\theta}^{a (1)}\partial_r \tilde{\theta}^{b (1)} +2 \psi \left[\gamma_0^{a c} \partial_c \tilde{\theta}^{b (1)}+ (a\leftrightarrow b) \right]+\gamma_0^{c d} \partial_c \tilde{\theta}^{a (1)}
 \partial_d \tilde{\theta}^{b (1)} \right.
 \nonumber \\
& &  \left.
 +\left[\gamma_0^{a c} \partial_c \tilde{\theta}^{b (2)}+ (a\leftrightarrow b) \right] \right\}.
\label{gammaab}
\eea

\section{Redshift and luminosity distance-redshift relation: going from geodesic light-cone to Poisson gauge}
\label{Sec3}
\setcounter{equation}{0}

\subsection{Redshift}
\label{Sec3A}

Let us begin with the redshift, starting from the non-perturbative solution (\ref{redshift}) and using the coordinate transformation 
defined above (in particular Eq.(\ref{Ups1})), we can obtain its second order perturbative expression in the PG for a general dark energy model with anisotropic stress. This was first done in \cite{BenDayan:2012wi} for the case with vanishing anisotropic case, but the final expression was not explicitly given. Here we present the final result in standard form, using the conformal time as affine parameter of our line-of-sight. To this aim we underline that since $\tau$  plays the role of the effective gauge-invariant velocity potential (see \cite{Fanizza:2013doa}), we can define in polar coordinates the spatial components of the perturbed velocity $v_\mu$ of the PG (geodesic) observer as:
\begin{equation}
v_i=(v_r+v_r^{(2)}, v_{\perp a}+v_{\perp a}^{(2)}) \quad \text{with} \quad v_r=-\partial_r \tau^{(1)}\quad,\quad v_r^{(2)}=-\partial_r \tau^{(2)} \quad,\quad v_{\perp a}=-\partial_a \tau^{(1)} \quad,\quad v_{\perp a}^{(2)}= -\partial_a \tau^{(2)}\,,
\label{GenVel}
\end{equation}
where $ \tau^{(1)}$ and $ \tau^{(2)}$ are the first- and second-order part of the coordinate transformation $\tau= \tau (\eta, r, \theta^a)$ between PG and GLC gauge (see Eq.(\ref{tau2order})). 
The unit vector $n_\mu$ along the direction connecting the source to the observer can be then expanded, 
in polar coordinates and to first order (which is enough for our purpose), as:
\begin{equation}
n^\mu=\left(0, -\frac{1}{a}(1+\psi), 0, 0\right)\,\,\,\,\,\,\,\,,\,\,\,\,\,\,\,\,n_\mu=\left(0, -a(1-\psi), 0, 0\right).
\end{equation}  
Taking then its scalar product with the spatial component of the perturbed velocity we have
\be 
\vec{v}\cdot \hat{n}=v_{||}+v_{||}^{(2)}=\partial_r P+ \psi \partial_r P +\frac{1}{2} \int_{\eta_{in}}^\eta d\eta' \frac{a(\eta')}{a(\eta)} \partial_r\left[ \phi^{(2)} - \phi^2 + ( \partial_r P )^2 + \gamma_0^{ab} ~ \partial_a P ~ \partial_b P \right] (\eta', r, \theta^a)
\label{vpar}
\ee
and we also have that
\be
v_{\perp a} v_{\perp}^a = \gamma_0^{ab} \partial_a P \partial_b P \,.
\label{vper}
\ee
Les us now define the following useful variables
\be
\psi^I=\frac{\psi+\phi}{2}\quad\quad\quad,\quad\quad\quad\psi^A=\frac{\psi-\phi}{2}\,,
\label{VariableIsAs}
\ee
which define the isotropic and anisotropic part of the Bardeen potential.
Hereafter we will use such variables to express our perturbed quantities.

Finally, also with the help of the following results
\begin{eqnarray}
Q_s &=& -2 \int_{\eta_s}^{\eta_o} d \eta' \psi^I\left(\eta', \eta_o-\eta', \theta^a_s\right) \label{Us1} 
\\
\partial_+ Q_s &=& \psi^I_o - \psi^I_s -2 \int_{\eta_s}^{\eta_o} d \eta' \partial_{\eta'}\psi^I\left(\eta', \eta_o-\eta', \theta^a_s\right) \label{Us2} 
\\
\partial_+\omega_s^{(2)} &=& \frac{1}{4}\left[\left(\phi_o^{(2)} + \psi_o^{(2)}\right)-\left(\phi_s^{(2)} + \psi_s^{(2)}\right)\right]
- \psi^I_s \left[ \psi^I_0 - \psi^I_s -2 \int_{\eta_s}^{\eta_o} d \eta' \partial_{\eta'}\psi^I\left(\eta', \eta_o-\eta', \theta^a_s\right) \right]
\nonumber \\ & & 
+2 \left(\psi_o^I \psi_o^A-\psi_s^I \psi_s^A\right) - \gamma_0^{ab} \partial_a \left( \int_{\eta_s}^{\eta_o} d \eta' \psi^I\left(\eta', \eta_o-\eta', \theta^a_s\right) \right) 
\partial_b \left( \int_{\eta_s}^{\eta_o} d \eta' \psi^I\left(\eta', \eta_o-\eta', \theta^a_s\right) \right)  \nonumber \\
&& 
-\frac{1}{2} \int_{\eta_s}^{\eta_0} d\eta' \partial_{\eta'}  \left[ \phi^{(2)} + \psi^{(2)} 
+8 \psi^I \psi^A
+ 4 \psi^I \left(\psi^I_o - \psi^I -2 \int_{\eta'}^{\eta_o} d \eta'' \partial_{\eta''}\psi^I\left(\eta'', \eta_o-\eta'', \theta^a_s\right)\right) \right.
\nonumber \\ & & \left.
+ 4 \gamma_0^{ab} \partial_a \left( \int_{\eta'}^{\eta_o} d \eta'' \psi^I\left(\eta'', \eta_o-\eta'', \theta^a_s\right) \right) 
\partial_b \left( \int_{\eta'}^{\eta_o} d \eta'' \psi^I\left(\eta'', \eta_o-\eta'', \theta^a_s\right) \right) \right] \left( \eta' , \eta_0 - \eta' , \theta^a_s \right)\,,
\label{Us3}
\end{eqnarray}
and 
\bea
& & \int_{\eta_s}^{\eta_0} d\eta' \partial_{\eta'}  \left[\psi^I\left(\eta'\right) \left(\psi^I_o - \psi^I\left(\eta'\right) -2 \int_{\eta'}^{\eta_o} d \eta'' \partial_{\eta''}\psi^I\left(\eta''\right)\right) 
+\gamma_0^{ab} \partial_a \left( \int_{\eta'}^{\eta_o} d \eta'' \psi^I\left(\eta''\right) \right) 
\partial_b \left( \int_{\eta'}^{\eta_o} d \eta'' \psi^I\left(\eta''\right) \right) \right]=
\nonumber \\
& & 
-\psi^I_o \int_{\eta_s}^{\eta_0} d\eta' \partial_{\eta'} \psi^I\left( \eta'\right)
-2\int_{\eta_s}^{\eta_0} d\eta' \left[
- \psi^I\left(\eta'\right) \partial_{\eta'}\psi^I\left(\eta'\right) 
- \partial_{\eta'}\psi^I\left(\eta'\right) \int_{\eta'}^{\eta_o} d \eta'' \partial_{\eta''}\psi^I\left(\eta''\right)
\right.
\nonumber \\ & & \left.
- \psi^I\left(\eta'\right) \int_{\eta'}^{\eta_o} d \eta'' \partial^2_{\eta''}\psi^I\left(\eta''\right)
+ \gamma_0^{ab} \partial_a \left( \int_{\eta'}^{\eta_o} d \eta'' \psi^I\left(\eta''\right) \right) 
\partial_b \left( \int_{\eta'}^{\eta_o} d \eta'' \partial_{\eta''}\psi^I\left(\eta''\right) \right) \right]\,,
\label{Us4}
\eea
we obtain the redshift up to second order in perturbation theory (hereafter we only use the conformal time $\eta$ as argument inside the integral over the line-of-sight, instead of extended arguments like $(\eta, \eta_o-\eta, \theta^a_s)$):
\be
1+z_s = \frac{a(\eta_o)}{a(\eta_s)} \left[1+ \delta^{(1)} z + \delta^{(2)} z \right]
\ee
with
\be
\delta^{(1)} z = v_{||o}- v_{||s}+(\psi^I_o-\psi^A_o)-(\psi^I_s-\psi^A_s)-2 \int_{\eta_s}^{\eta_o} d \eta' \partial_{\eta'}\psi^I\left(\eta'\right)
\label{Redshift1}
\ee
\begin{eqnarray}
\delta^{(2)} z &=& v^{(2)}_{||o}- v^{(2)}_{||s}+\frac{1}{2}\left(\phi_o^{(2)}-\phi_s^{(2)}\right)
-\frac{1}{2} \int_{\eta_s}^{\eta_0} d\eta' \partial_{\eta'}  \left[ \phi^{(2)}\left( \eta'\right) + \psi^{(2)}\left( \eta'\right)\right]
+\frac{1}{2}\left(v_{||o}- v_{||s}\right)^2
+\frac{1}{2}\left((\psi^I_s)^2-(\psi^I_o)^2\right)\nonumber \\ & & 
+\left(v_{||o}- v_{||s}-\psi^I_s\right)\left(\psi^I_o-\psi^I_s-2 \int_{\eta_s}^{\eta_o} d \eta' \partial_{\eta'}\psi^I\left(\eta'\right)\right)
+\frac{1}{2} \left( v^a_{\perp s} v_{\perp a\,s} 
-  v^a_{\perp o} v_{\perp a\,o}\right) 
-2 a \, v^a_{\perp s} \partial_a \int_{\eta_s}^{\eta_o} d \eta' \psi^I\left(\eta'\right)
\nonumber \\ & &
-2 \psi^I_o \int_{\eta_s}^{\eta_0} d\eta' \partial_{\eta'} \psi^I\left( \eta'\right)
+4\int_{\eta_s}^{\eta_0} d\eta' \left[\psi^I\left(\eta'\right) \partial_{\eta'}\psi^I\left(\eta'\right) 
+ \partial_{\eta'}\psi^I\left(\eta'\right) \int_{\eta'}^{\eta_o} d \eta'' \partial_{\eta''}\psi^I\left(\eta''\right)
\right.
\nonumber \\ & & \left.
+\psi^I\left(\eta'\right) \int_{\eta'}^{\eta_o} d \eta'' \partial^2_{\eta''}\psi^I\left(\eta''\right)
- \gamma_0^{ab} \partial_a \left( \int_{\eta'}^{\eta_o} d \eta'' \psi^I\left(\eta''\right) \right) 
\partial_b \left( \int_{\eta'}^{\eta_o} d \eta'' \partial_{\eta''}\psi^I\left(\eta''\right) \right) \right]
\nonumber \\ & &
+\left(\psi^A_s-\psi^A_o\right) \left(v_{||o}- v_{||s}\right)+\frac{3}{2} (\psi^A_s)^2-\frac{1}{2} (\psi^A_o)^2
-\psi^A_o \psi^A_s-3\psi^I_s\psi^A_s+\psi^I_o\psi^A_o+\psi^A_s\psi^I_o+\psi^A_o\psi^I_s
\nonumber \\ & &
-4 \int_{\eta_s}^{\eta_o} d \eta' \partial_{\eta'}(\psi^I\psi^A)\left(\eta'\right)
-2 \left(\psi^A_s-\psi^A_o\right) \int_{\eta_s}^{\eta_0} d\eta' \partial_{\eta'} \psi^I\left( \eta'\right)
\label{Redshift2}
\end{eqnarray}
Let us underline how the results above are still written in terms of the angles of the source position. 

On the other hand,  the perturbed redshift should be written as a function 
of the observer's angular coordinates.
Using the properties of the GLC gauge, this corresponds to writing the redshift as a function of the GLC angular coordinates $\tilde{\theta}^a$.
In fact, as recalled in the previous section, $\tilde{\theta}^a$ are equivalent to the standard angular coordinate at the observer position and are constant along the line-of-sight. Therefore the perturbed redshift $1+\bar{z}_s$ is given by Taylor-expanding $1+z_s$ around $\tilde{\theta}^a$ (we use a bar to denote that the redshift is now expressed in terms of 
$\tilde{\theta}^a$). To this purpose it is enough to invert Eq.(\ref{thetatilde2orderShort}) to first order since the background redshift is independent from the angles, therefore we need the expansion
\be
{\theta}^a = {\theta}^{a (0)}+{\theta}^{a (1)}
= \tilde{\theta}^a - 2 \int_{\eta_s^{(0)}}^{\eta_o} d\eta'~ \gamma_0^{ab} \partial_b \int_{\eta'}^{\eta_o} d\eta''~ \psi^I(\eta'')\,.
\label{thetatilde1orderShort_Inverted}
\ee
Then the redshift as function of the observer's angular coordinates $\tilde{\theta}^a$ will be given by Eqs. (\ref{Redshift1}) and (\ref{Redshift2}) with $\theta^a_s$ replaced by $\tilde{\theta}^a$ plus a further term given by Taylor-expanding $\delta^{(1)} z $ around $\tilde{\theta}^a$. 
Namely we have
\be
\delta^{(1)} \bar{z} = \delta^{(1)} z|_{\theta^a_s=\tilde{\theta}^a}
\label{Redshift1bar}
\ee
\bea
\delta^{(2)} \bar{z} &=& \delta^{(2)} z|_{\theta^a_s=\tilde{\theta}^a}+2 \partial_a  \left(v_{||s}+\psi^I_s-\psi^A_s\right)  \int_{\eta_s}^{\eta_o}\!d \eta' \gamma_0^{ab} \partial_b \int_{\eta'}^{\eta_o}\!d \eta'' \psi^I\left(\eta''\right)
\nonumber \\
& & 
+4 \int_{\eta_s}^{\eta_o}\!\!d\!\eta'\partial_a \left(\partial_{\eta'}\psi^I\left(\eta'\right)\right)
\!\int_{\eta_s}^{\eta_o}\!\!\!d\!\eta''\gamma_0^{ab} \partial_b\!\int_{\eta''}^{\eta_o}\!d\!\eta'''\psi^I\left(\eta'''\right)
\label{Redshift2bar}
\eea
Let us point out that, in particular, the perturbed redshift, given in Eqs. (\ref{Redshift1}), (\ref{Redshift2}), (\ref{Redshift1bar}) and (\ref{Redshift2bar}), is uniquely defined after we go to the fully gauge fixed  PG, even if we start from the GLC gauge where some residual gauge freedom is stil present \cite{Fanizza:2013doa}.

\subsection{Luminosity Distance}

Let us now move to the luminosity distance $d_L$. We first give the basic steps to arrive at the final expression for the luminosity distance-redshift relation to a given order in perturbation theory, in the case under consideration the second, without going into full details (see \cite{BenDayan:2012wi,Fanizza:2013doa}). 
The first step consists in writing $d_L$ up to second order in perturbation theory in PG using the exact expression in Eq.(\ref{lumdist}) and the second order coordinate transformation given in Eqs (\ref{tau2order})-(\ref{gammaab}).
On the other hand, as mentioned, we want to write $d_L$ as function of the observed redshift and with respect to the angles at the observer position. We can first write $d_L$ as function of the observed redshift defining a fiducial model with coordinates 
$(\eta_s^{(0)}, r_s^{(0)}, \theta^a_s)$ for which the observed redshift and the past light-cone of our observer are given by
\be 
1+z_s=\frac{a(\eta_o)}{a(\eta_s^{(0)})}  \,\,\,\,\,\,\,\,\,\,\,\,\,\,\,\,\,\,,\,\,\,\,\,\,\,\,\,\,\,\,\,\,\,\,\,\, w=w_o=\eta_o=\eta_s^{(0)}+r_s^{(0)}  \,.
\ee
We then expand conformal time and radial PG coordinates around the coordinates of the fiducial model as $\eta_s=\eta_s^{(0)}+\eta_s^{(1)}
+\eta_s^{(2)}$ and $r_s=r_s^{(0)}+r_s^{(1)}+r_s^{(2)}$, and obtain the terms of these expansions by 
perturbatively solving the following system of 
equations:
\begin{eqnarray}
& & 1+z_s=\frac{a(\eta_o)}{a(\eta_s^{(0)})}=\frac{a(\eta_o)}{a(\eta_s)} \left[1+ \delta^{(1)} z + \delta^{(2)} z \right] \\
& & w = \eta_o = w^{(0)}+w^{(1)}+w^{(2)}
\end{eqnarray}
where we have to use Eqs. (\ref{w2order}), (\ref{Redshift1}) and (\ref{Redshift2}).
In particular, for the case under consideration we have the following solution for our fiducial model
\be 
\eta_s^{(1)}=\frac{ \delta^{(1)} z}{\Hcal_s}
\ee
\be
\eta_s^{(2)}=\frac{1}{\Hcal_s}\left[\delta^{(2)} z+\delta^{(1\rightarrow 2)} z-\frac{1}{2}\left(1+\frac{\Hcal_s'}{\Hcal_s^2}\right)(\delta^{(1)} z)^2\right]
\ee
and 
\be 
r_s^{(0)}=\eta_o-\eta_s^{(0)}=\Delta\eta \quad\quad\quad\quad,\quad\quad\quad\quad r_s^{(1)}=-\eta_s^{(1)}+2 
 \int_{\eta_s^{(0)}}^{\eta_o} d \eta' \psi^I\left(\eta'\right)
\ee
\be
r_s^{(2)}=-\eta_s^{(2)}-w_s^{(2)}-w_s^{(1\rightarrow 2)}
\ee
where $\Hcal_s=a'(\eta_s^{(0)})/a(\eta_s^{(0)})$ is the comoving Hubble parameter of the fiducial model, 
and the quantities $\delta^{(1\rightarrow 2)} z$ and $w_s^{(1\rightarrow 2)}$ stand for the second order contribution coming from Taylor expanding $\delta^{(1)} z$ and $w_s^{(1)}$ around the background source position of our fiducial model. These are given by 
\bea 
\delta^{(1\rightarrow 2)} z
&=& \eta_s^{(1)}\left( \partial_\eta \psi^I_s+\partial_\eta \psi^A_s+\Hcal_s v_{||s}
+\partial_r v_{||s}\right)+\left[\frac{3}{2} \partial_\eta \psi^I_o+\frac{1}{2} \partial_r \psi^I_o-2 \partial_\eta \psi^I_s-\partial_r \psi^I_s+\partial_r \psi^A_s
\right. \nonumber \\
 & & \left.
-\partial_r v_{||s}-2 \int_{\eta^{(0)}_s}^{\eta_o} d\eta'
\partial_{\eta'}^2 \psi^I(\eta')\right] \left(2 \int_{\eta^{(0)}_s}^{\eta_o} d\eta' \psi^I(\eta')\right)
\eea
\bea
w^{(1\rightarrow 2)}
&=&\left[\psi^I_o-\psi^I_s
-2 \int_{\eta_s^{(0)}}^{\eta_o} d \eta' \partial_{\eta'}\psi^I\left(\eta'\right)\right]\left(2 \int_{\eta^{(0)}_s}^{\eta_o} d\eta' \psi^I(\eta')\right) +\psi_s^I \left[2 \eta_s^{(1)}-2 \int_{\eta^{(0)}_s}^{\eta_o} d\eta' \psi^I(\eta')\right]\,.
\eea
where, we underline, all the quantities above are now expressed with respect to the background conformal time of our fiducial model. 

Once we have $(\eta_s^{(0)}, r_s^{(0)})$, $(\eta_s^{(1)}, r_s^{(1)})$ and $(\eta_s^{(2)}, r_s^{(2)})$ we can obtain the luminosity 
distance-redshift relation $d_L(z_s, \theta^a_s)$ by Taylor expanding the second order solution for $d_L$, previously found as function of the PG coordinates, around the fiducial values $(\eta_s^{(0)}, r_s^{(0)})$.

This yields $d_L$ as function of $\eta_s^{(0)}$, which determines the observed redshift. The last step is to write it as function 
of the observer's angular coordinates.
We proceed as done for the observed redshift, using the properties of the GLC gauge for which $\tilde{\theta}^a$ are equivalent to the standard angular coordinate at the observer position and are constant along the line-of-sight. Therefore the luminosity distance $\bar{d}_L(z_s, \tilde{\theta}^a)$ is given by Taylor expanding $d_L(z_s, {\theta}^a_s)$ around $\tilde{\theta}^a$ (we use a bar to denote that the luminosity distance is now expressed in terms of 
$\tilde{\theta}^a$). To this purpose it is enough to use Eq.(\ref{thetatilde1orderShort_Inverted}), as for the redshift the background value $d_L^{(0)}$ is independent from the angles.

In \cite{BenDayan:2012wi} the procedure above is followed in full details for the case with vanishing anisotropic stress.
Here, apart from the above results, we give only the final results, skipping the major part of the technical details.
Therefore, following the procedure summarized above,  we obtain
\beq
\frac{\bar{d}_L(z_s, \tilde{\theta}^a)}{(1+z_s)a_o \Delta \eta}
= {\bar{d}_L(z_s, \tilde{\theta}^a)\over d_L^{FLRW}(z_s)} =  1 + \bar{\delta}_S^{(1)}(z_s, \tilde{\theta}^a) + \bar{\delta}_S^{(2)}(z_s, \tilde{\theta}^a) ~~,
\eeq
where the first order luminosity distance is given by
\bea
\bar{\delta}_S^{(1)}(z_s, \tilde{\theta}^a) &=& -\left(1-\frac{1}{\Hcal_s\Delta \eta}\right) v_{||s}-\frac{1}{\Hcal_s\Delta \eta} v_{||o}
-(\psi^I_s+\psi_s^A)+
\left(1-\frac{1}{\Hcal_s\Delta \eta}\right)\left[(\psi^I_o-\psi^A_o)-(\psi^I_s-\psi_s^A)
\right.
\nonumber \\ & &
\left.
-2 \int_{\eta_s^{(0)}}^{\eta_o} d \eta' \partial_{\eta'}\psi^I\left(\eta'\right)\right]
+\frac{2}{\Delta \eta}  \int_{\eta^{(0)}_s}^{\eta_o} d\eta' \psi^I(\eta')-\frac{1}{\Delta\eta} \int_{\eta_s^{(0)}}^{\eta_o} d \eta' \,\frac {\eta' - \eta_s^{(0)}}{\eta_o - \eta'} \Delta_2 \psi^I(\eta')
\label{finaldLord1}
\eea
with $\Delta_2=\partial^2_\theta+\cot \theta \partial_\theta +1/(\sin \theta)^2 \partial_\phi^2$ the 2-dimensional angular Laplacian.
This is in full agreement with the previous results of \cite{Bonvin:2005ps,Pyne:2003bn} and with the ones of \cite{Fanizza:2013doa} for the case of vanishing anisotropic case.

The second order result is much more involved.
We split it in three different parts: 
\be
\bar{\delta}_S^{(2)}(z_s, \tilde{\theta}^a) = \bar{\delta}_{path}^{(2)} +  \bar{\delta}_{pos}^{(2)} +  \bar{\delta}_{mixed}^{(2)} \,,
\label{finaldLord2} 
\ee
where $\bar{\delta}_{path}^{(2)}$ denotes terms connected to the photon path and to the boundary terms; $\bar{\delta}_{pos}^{(2)}$ is for the terms manifestly generated by the source and observer peculiar velocity. Finally, $\bar{\delta}_{mixed}^{(2)}$ mixes peculiar velocity effects with all others. 
We then obtain the following final result for $\bar{\delta}_{pos}^{(2)}$
\bea
\bar{\delta}_{pos}^{(2)} &=& \left(1-\frac{1}{ \Hcal_s \Delta \eta}\right) \left[\frac{1}{2} v^a_{\perp s} v_{\perp a\, s}-\frac{1}{ \Hcal_s}
\left(v_{||s}-v_{||o}\right) \partial_r v_{||s} - v^{(2)}_{||s}\right]
 \nonumber \\ 
& & 
+\frac{1}{ \Hcal_s \Delta \eta} \left(\frac{1}{2} v^a_{\perp o} v_{\perp a\, o}- v^{(2)}_{||o}\right)
-\frac{1}{2}v^{2}_{||s}+v_{||o}v_{||s}+\frac{1}{2}\frac{1}{ \Hcal_s \Delta \eta} \frac{\Hcal_s'}{\Hcal^2_s}\left(v_{||s}-v_{||o}\right)^2
\label{posAS}
\eea
For the parts $\bar{\delta}_{mixed}^{(2)}$ and $\bar{\delta}_{path}^{(2)}$ we perform a further split:
$\bar{\delta}_{mixed,a}^{(2)}$ and $\bar{\delta}_{path,a}^{(2)}$ contain terms which depend only on 
 $\psi^I$ (and, in case, from the genuine second order variables), 
while $\bar{\delta}_{mixed,b}^{(2)}$ and $\bar{\delta}_{path,b}^{(2)}$ contain the rest of the terms which depend also on $\psi^A$.
We then have 
\bea
\bar{\delta}_{mixed,a}^{(2)} &=&  \left(1-\frac{1}{ \Hcal_s \Delta \eta}\right)\left[+ \psi^I_s  v_{||s}+v_{||s} 
\frac{1}{\Delta \eta} \int_{\eta_s}^{\eta_o} d \eta' \frac {\eta' - \eta_s}{\eta_o - \eta'} \Delta_2 \psi^I\left(\eta'\right)
-2 a v^a_{\perp s} \partial_a \int_{\eta_s}^{\eta_o}\!d \eta'\psi^I\left(\eta'\right)
\right. \nonumber \\ 
& & \left.
+\frac{1}{ \Hcal_s} \left(\psi^I_o-\psi^I_s-2\int_{\eta_s}^{\eta_o}\!d \eta' \partial_{\eta'}\psi^I\left(\eta'\right)
-2\Hcal_s \int_{\eta_s}^{\eta_o}\!d \eta' \psi^I\left(\eta'\right)\right)\partial_r v_{||s} 
\right. \nonumber \\ 
& & \left.
+2 \partial_a v_{||s} \int_{\eta_s}^{\eta_o}\!d \eta' \gamma_0^{ab} \partial_b \int_{\eta'}^{\eta_o}\!d \eta'' \psi^I\left(\eta''\right)\right]
+v_{||o}\frac{1}{ \Hcal_s \Delta \eta}\left(\psi^I_s+\frac{1}{\Delta \eta} \int_{\eta_s}^{\eta_o} d \eta' \frac {\eta' - \eta_s}{\eta_o - \eta'} \Delta_2 \psi^I\left(\eta'\right)\right)\nonumber \\ 
& &
-v_{||s} \frac{2}{\Delta \eta} \int_{\eta_s}^{\eta_o} d \eta' \psi^I\left(\eta'\right)
+\left(v_{||s}-v_{||o}\right)\left[\frac{1}{ \Hcal_s \Delta \eta}\left(1-\frac{\Hcal_s'}{\Hcal^2_s}\right)\left(\psi^I_o-\psi^I_s-2\int_{\eta_s}^{\eta_o}\!d \eta' \partial_{\eta'}\psi^I\left(\eta'\right)\right)
\right. \nonumber \\ 
& & \left.
+ \frac{2}{\Hcal_s \Delta \eta} \psi^I_s+\frac{1}{\Hcal^2_s \Delta \eta} \partial_\eta \psi^I_s-\frac{1}{\Hcal_s} \partial_r \psi^I_s
-\frac{1}{\Hcal_s \Delta \eta^2}\int_{\eta_s}^{\eta_o}\!d \eta' \Delta_2\psi^I\left(\eta'\right)
\right]
\nonumber 
\\
& & 
{-\frac{1}{ \Hcal_s} \partial_a \left(v_{||s}-v_{||o}\right)\gamma_{0s}^{ab}  \int_{\eta_s}^{\eta_o}\!\!\!d\eta' \partial_b \psi^I(\eta')}
\label{mixedASa}
\eea

\bea
\bar{\delta}_{mixed,b}^{(2)} &=&  \left(1-\frac{1}{ \Hcal_s \Delta \eta}\right)\left[
+\psi^A_s  v_{||s}+
\frac{1}{ \Hcal_s} \partial_r v_{||s} (\psi^A_s-\psi^A_o)\right]
-\frac{1}{ \Hcal_s \Delta \eta} \frac{\Hcal_s'}{\Hcal^2_s}(\psi^A_s-\psi^A_o)
 \left(v_{||s}-v_{||o}\right)
  \nonumber \\ 
& & 
+\frac{1}{ \Hcal_s \Delta \eta}(2 \psi^A_s-\psi^A_o)v_{||s}
{+\left(\frac{1}{ \Hcal_s \Delta \eta}  \partial_\eta \psi^A_s-\partial_r \psi^A_s\right) \frac{1}{ \Hcal_s} \left(v_{||s}-v_{||o}\right)}
 \label{mixedASb}
\eea

and
\bea
& & \!\!\!\!\!\bar{\delta}_{path,a}^{(2)}\!=\!\left(1-\frac{1}{ \Hcal_s \Delta \eta}\right) \left\{
\frac{1}{2}\left(\phi_o^{(2)}-\phi_s^{(2)}\right)-\frac{1}{2}\int_{\eta_s}^{\eta_o}\!d \eta' \partial_{\eta'}\left(
\psi^{(2)}\left(\eta'\right)+\phi^{(2)}\left(\eta'\right)\right)\right\}-\frac{1}{2}\psi_s^{(2)}
\nonumber \\ 
& &
-\frac{1}{4} \frac{1}{\Delta \eta}\int_{\eta_s}^{\eta_o}\!d \eta' \frac {\eta' - \eta_s}{\eta_o - \eta'} \Delta_2\left(
\psi^{(2)}\left(\eta'\right)+\phi^{(2)}\left(\eta'\right)\right)
+\frac{1}{2} \frac{1}{\Delta \eta}\int_{\eta_s}^{\eta_o}\!d \eta'\left(
\psi^{(2)}\left(\eta'\right)+\phi^{(2)}\left(\eta'\right)\right)
\nonumber \\ 
& &
+ \left(1-\frac{1}{ \Hcal_s \Delta \eta}\right) \left\{
\left[-\frac{1}{2}\partial_r\psi^I_o-\frac{3}{2}\partial_\eta\psi^I_o+\partial_r\psi^I_s
+2 \partial_\eta\psi^I_s+2 \int_{\eta_s}^{\eta_o}\!d \eta' \partial^2_{\eta'}\psi^I\left(\eta'\right)\right]
\left(-2 \int_{\eta_s}^{\eta_o}\!d \eta' \psi^I\left(\eta'\right)\right)
\right. \nonumber \\ 
& & \left.
-\left(\psi^I_o-\psi^I_s-2 \int_{\eta_s}^{\eta_o}\!d \eta' \partial_{\eta'}\psi^I\left(\eta'\right)\right)
\frac{1}{\Delta \eta}\int_{\eta_s}^{\eta_o}\!d \eta' \frac {\eta' - \eta_s}{\eta_o - \eta'} \Delta_2
\psi^I\left(\eta'\right)+\frac{1}{2}(\psi^I_s)^2-\frac{1}{2}(\psi^I_o)^2
\right.\nonumber \\ 
& & \left.
-2 \psi^I_o \int_{\eta_s}^{\eta_0} d\eta' \partial_{\eta'} \psi^I\left( \eta'\right)
-4\int_{\eta_s}^{\eta_0} d\eta' \left[
- \psi^I\left(\eta'\right) \partial_{\eta'}\psi^I\left(\eta'\right) 
- \partial_{\eta'}\psi^I\left(\eta'\right) \int_{\eta'}^{\eta_o} d \eta'' \partial_{\eta''}\psi^I\left(\eta''\right)
\right.\right.
\nonumber \\ 
& & \left.\left.
- \psi^I\left(\eta'\right) \int_{\eta'}^{\eta_o} d \eta'' \partial^2_{\eta''}\psi^I\left(\eta''\right)
+ \gamma_0^{ab}\partial_a \left( \int_{\eta'}^{\eta_o} d \eta'' \psi^I\left(\eta''\right) \right) 
\partial_b \left( \int_{\eta'}^{\eta_o} d \eta'' \partial_{\eta''}\psi^I\left(\eta''\right) \right) \right]
\right. \nonumber \\ 
& & \left.
\!+\!2\partial_a\! \psi^I_s
\!\int_{\eta_s}^{\eta_o}\!\!\!d\!\eta'\gamma_0^{ab} \partial_b\!\int_{\eta'}^{\eta_o}\!d\!\eta''\psi^I\left(\eta''\right)
+4 \int_{\eta_s}^{\eta_o}\!\!d\!\eta'\partial_a \left(\partial_{\eta'}\psi^I\left(\eta'\right)\right)
\!\int_{\eta_s}^{\eta_o}\!\!\!d\!\eta''\gamma_0^{ab} \partial_b\!\int_{\eta''}^{\eta_o}\!d\!\eta'''\psi^I\left(\eta'''\right)
\right\}
+4 \psi^I_s \int_{\eta_s}^{\eta_o}\!d \eta' \partial_{\eta'}\psi^I\left(\eta'\right)
\nonumber \\ 
& &
+\frac{3}{2}(\psi^I_s)^2-2 \psi^I_s \psi^I_o+\frac{1}{\Hcal_s}\left(\partial_r \psi^I_s-\frac{1}{ \Hcal_s \Delta \eta}\partial_\eta \psi^I_s\right)\left(\psi^I_o-\psi^I_s-2 \int_{\eta_s}^{\eta_o}\!d \eta' \partial_{\eta'}\psi^I\left(\eta'\right)\right)
-2\partial_r \psi^I_s\int_{\eta_s}^{\eta_o}\!d \eta'\psi^I\left(\eta'\right)
\nonumber \\ 
& &
-\frac{1}{2}\frac{1}{ \Hcal_s \Delta \eta}\left(1-\frac{\Hcal_s'}{\Hcal^2_s}\right)
\left[\left(\psi^I_s-\psi^I_o\right)^2+2\left(\psi^I_s-\psi^I_o\right)\left(2 \int_{\eta_s}^{\eta_o}\!d \eta' \partial_{\eta'}\psi^I\left(\eta'\right)\right)+4 \left(\int_{\eta_s}^{\eta_o}\!d \eta' \partial_{\eta'}\psi^I\left(\eta'\right)\right)^2\right]
\nonumber \\
& &
+\frac{2}{\Delta \eta}\int_{\eta_s}^{\eta_o}\!d \eta'\!\left[\psi^I\left(\eta'\right)\left(\psi^I_o-\psi^I\left(\eta'\right) 
-2 \int_{\eta'}^{\eta_o}\!d\!\eta'' \partial_{\eta''}\psi^I\left(\eta''\right)\right)
+\gamma_0^{ab}\partial_a\!\left(\int_{\eta'}^{\eta_o}\!\!\!d\!\eta''\!\psi^I\left(\eta''\right)\right)\!\partial_b\!\left(\int_{\eta'}^{\eta_o}\!\!\!d\!\eta''\!\psi^I\left(\eta''\right)\right)\right] 
\nonumber 
\\ 
& &
+\left(\psi^I_s-\frac{2}{\Delta \eta}\int_{\eta_s}^{\eta_o}\!d \eta'\psi^I\left(\eta'\right)\right)
\frac{1}{\Delta \eta}\int_{\eta_s}^{\eta_o}\!d \eta' \frac {\eta' - \eta_s}{\eta_o - \eta'} \Delta_2\psi^I\left(\eta'\right)
+\frac{1}{2}\left(\frac{1}{\Delta \eta}\int_{\eta_s}^{\eta_o}\!d \eta' \frac {\eta' - \eta_s}{\eta_o - \eta'} \Delta_2\psi^I\left(\eta'\right)\right)^2
\nonumber 
\eea
\bea
& &
+\left[\frac{1}{ \Hcal_s \Delta \eta}\left(\psi^I_o-\psi^I_s-2 \int_{\eta_s}^{\eta_o}\!d \eta' \partial_{\eta'}\psi^I\left(\eta'\right)\right)
-\frac{1}{\Delta \eta}\int_{\eta_s}^{\eta_o}\!d \eta'\psi^I\left(\eta'\right)\right]\frac{1}{\Delta \eta}\int_{\eta_s}^{\eta_o}\!d \eta' \Delta_2\psi^I\left(\eta'\right)
\nonumber \\ 
& &
-\left[ \int_{\eta_s}^{\eta_o}\!\!\!d\eta' \frac{1}{(\eta_o-\eta')^2}\left(\psi^I(\eta')-\psi^I_o\right)+2
 \int_{\eta_s}^{\eta_o}\!\!\!d\eta'  \frac{1}{(\eta_o-\eta')^2} \int_{\eta'}^{\eta_o}\!\!\!d\eta''\partial_{\eta''}\psi^I(\eta'')\right]
  \int_{\eta_s}^{\eta_o}\!\!\!d\eta' \Delta_2 \psi^I(\eta')
\nonumber \\ 
& &
+\!2\partial_a\!\psi^I_s
\!\int_{\eta_s}^{\eta_o}\!\!\!d\!\eta'\gamma_0^{ab}\partial_b\!\int_{\eta'}^{\eta_o}\!d\!\eta''\psi^I\left(\eta''\right)
-\frac{4}{\Delta \eta}\left[\int_{\eta_s}^{\eta_o}\!d \eta' \partial_a \psi^I\left(\eta'\right)
\!\int_{\eta_s}^{\eta_o}\!\!\!d\!\eta''\gamma_0^{ab}\partial_b\!\int_{\eta''}^{\eta_o}\!d\!\eta'''\psi^I\left(\eta'''\right)\right]
\nonumber
\\
& &
+\left(\partial_a\!\int_{\eta_s}^{\eta_o}\!d\!\eta'\psi^I\left(\eta'\right)\right)
\left[4 \int_{\eta_s}^{\eta_o}\!\!\!d\eta' \frac{1}{(\eta_o-\eta')}\gamma_0^{ab}
\int_{\eta'}^{\eta_o}\!\!\!d\eta'' \partial_b \psi^I(\eta'') - 3 \int_{\eta_s}^{\eta_o}\!\!\!d\eta' \gamma_0^{ab}\partial_b\psi^I(\eta')
\right.
\nonumber \\ 
& &
\left.
-6
 \int_{\eta_s}^{\eta_o}\!\!\!d\eta' \gamma_0^{ab}
\int_{\eta'}^{\eta_o}\!\!\!d\eta'' \partial_b\partial_{\eta''}\psi^I(\eta'')
\right]
+\partial_a\left(\int_{\eta_s}^{\eta_o}\!\!\!d\!\eta'\gamma_0^{bd}\partial_d\!\int_{\eta'}^{\eta_o}\!d\!\eta''\psi^I\left(\eta''\right)\right)
\partial_b\left(\int_{\eta_s}^{\eta_o}\!\!\!d\!\eta'\gamma_0^{ac}\partial_c\!\int_{\eta'}^{\eta_o}\!d\!\eta''\psi^I\left(\eta''\right)\right)
\nonumber \\
& &
-2\left(\!\int_{\eta_s}^{\eta_o}\!\!\!d \eta'\psi^I\left(\eta'\right)\right)\!\int_{\eta_s}^{\eta_o}\!\!\!d \eta'\left[-\frac{1}{(\eta_o-\eta')^3}
\!\int_{\eta'}^{\eta_o}\!\!\!d\!\eta''\Delta_2\psi^I\left(\eta''\right)
+\frac{1}{(\eta_o-\eta')^2}\!\left(\frac{1}{2} \Delta_2\psi^I\left(\eta'\right)+\int_{\eta'}^{\eta_o}\!\!\!d\!\eta''\partial_{\eta''}\left(\Delta_2\psi^I\left(\eta''\right)\right)
\right)
\right]
\nonumber \\ 
& &
+\partial_a \left\{\frac{1}{ \Hcal_s} \left[-\psi^I_s-2\int_{\eta_s}^{\eta_o}\!\!\!d\!\eta'\!\partial_{\eta'}\psi^I\left(\eta'\right)\right]
- \int_{\eta_s}^{\eta_o}\!\!\!d\eta'\psi^I(\eta')\right\}\gamma_{0s}^{ab}  \int_{\eta_s}^{\eta_o}\!\!\!d\eta' \partial_b \psi^I(\eta')
\nonumber \\ 
& &
+\frac{2}{\eta_o-\eta_s} \int_{\eta_s}^{\eta_o}\!\!\!d\eta'  \frac{\eta'-\eta_s}{\eta_o-\eta'} \partial_b 
\left[ \Delta_2 \psi^I(\eta')\right]
\int_{\eta_s}^{\eta_o} d \eta'' 
\gamma_0^{ab} \partial_a \int_{\eta''}^{\eta_o}\!d \eta'''  \psi^I(\eta''')
\nonumber
\\ 
& &
+\frac{1}{\left(\sin\tilde{\theta}\right)^2}\left[\frac{1}{\Delta \eta}\int_{\eta_s}^{\eta_o}\!\!\!d \eta' \frac{\eta'-\eta_s}{\eta_o-\eta'}
\partial_{\tilde{\theta}}\psi^I\left(\eta'\right)\right]^2
-\frac{1}{\Delta \eta} \int_{\eta_s}^{\eta_o}\!\!\!d\!\eta' \frac{\eta'-\eta_s}{\eta_o - \eta'} \Delta_2 \left[
\psi^I\left(\eta'\right) \left(\psi^I_o-\psi^I\left(\eta'\right)-2\int_{\eta'}^{\eta_o}\!\!\!d\!\eta''\!\partial_{\eta''}\psi^I\left(\eta''\right)\right)
\right.
\nonumber \\ 
& & \left.
+\gamma_0^{ab}\partial_a\left( \int_{\eta'}^{\eta_o}\!\!\!d \eta'' \psi^I\left(\eta''\right) \right) 
\partial_b \left( \int_{\eta'}^{\eta_o} d \eta'' \psi^I\left(\eta''\right) \right) \right]{-} \int_{\eta_s}^{\eta_o} d \eta'\left\{ 
\psi^I\left(\eta'\right) \lim_{\bar{\eta}\rightarrow \eta_o}\left(\frac{1}{(\eta_o-\bar{\eta})^2}  \int_{\bar{\eta}}^{\eta_o} d \eta'' \Delta_2
\psi^I\left(\eta''\right)\right) 
\right. \nonumber \\ 
& & \left. 
-2 \psi^I\left(\eta'\right)\frac{1}{\eta_o-\eta'}\int_{\eta'}^{\eta_o} d \eta''  \frac {\eta'' - \eta'}{\eta_o - \eta''}\Delta_2 \partial_{\eta''} \psi^I\left(\eta''\right)
+2 \gamma_0^{ab}\partial_b \left(\int_{\eta'}^{\eta_o} d \eta'' \psi^I\left(\eta''\right)\right) \frac{1}{\eta_o-\eta'}\!\int_{\eta'}^{\eta_o}\!d \eta''  \frac {\eta''-\eta'}{\eta_o-\eta''}\partial_a \Delta_2 \psi^I\left(\eta''\right)
\right. \nonumber \\ 
& & \left.
-\left(\!\psi^I_o-2\psi^I\left(\eta'\right)-2\int_{\eta'}^{\eta_o}\!\!\!d\!\eta''\!\partial_{\eta''}\psi^I\left(\eta''\right)\right)
\frac{1}{(\eta_o-\eta')^2}\int_{\eta'}^{\eta_o}\!\!\!d\!\eta''\!\Delta_2 \psi^I\left(\eta''\right)
+\partial_a \psi\left(\eta'\right)\left[
 \lim_{\bar{\eta}\rightarrow \eta_o}\left(\gamma_0^{ab} \partial_b\!\!\int_{\bar{\eta}}^{\eta_o} d \eta''
\psi\left(\eta''\right)\right) 
\right. \right. \nonumber \\ 
& & \left. \left.
-2 \int_{\eta'}^{\eta_o}\!d \eta'' \gamma_0^{ab} \partial_b \int_{\eta''}^{\eta_o}\!d \eta'''\partial_{\eta'''}\psi\left(\eta'''\right) 
\right]
+2 \partial_a\!\!\left[\gamma_0^{db}\partial_b\!\!\int_{\eta'}^{\eta_o}\!\!\!d\!\eta''\!\psi^I\left(\eta''\right)\right]\int_{\eta'}^{\eta_o}\!\!\!d\!\eta''\! \partial_d
\left[\gamma_0^{ac}\partial_c\!\!\int_{\eta''}^{\eta_o}\!\!\!d\!\eta'''\!\psi^I\left(\eta'''\right)\right]
\right. \nonumber \\ 
& & \left.
+2\gamma_0^{ab}\partial_a\!\!\left(\psi^I\left(\eta'\right)+\int_{\eta'}^{\eta_o}\!\!\!d\!\eta''\!\partial_{\eta''}\psi^I\left(\eta''\right)\right)
\partial_b\!\!\int_{\eta'}^{\eta_o}\!\!\!d\eta''\!\psi^I\left(\eta''\right)
\right\}
\label{pathASa}
\eea

\bea
\!\!\!\!\!\bar{\delta}_{path,b}^{(2)}\!&=&\!\left(1-\frac{1}{ \Hcal_s \Delta \eta}\right) \left\{
-(\psi^A_s- \psi^A_o)\frac{1}{\Delta \eta}\int_{\eta_s}^{\eta_o} d \eta'  \frac {\eta' - \eta_s}{\eta_o - \eta'}\Delta_2  \psi^I\left(\eta'\right)
-\psi^A_s \psi^A_o+(\psi^A_s)^2+\psi^A_s \psi^I_o
+\psi^A_o \psi^I_s-2 \psi^A_s \psi^I_s
\right. \nonumber \\
& & \left.
-4 \int_{\eta_s}^{\eta_o} d \eta' \partial_{\eta'} (\psi^I\psi^A)\left(\eta'\right)
\right\}
{-\left(\frac{1}{ \Hcal_s \Delta \eta}\partial_\eta  \psi^I_s-\partial_r  \psi^I_s\right)\frac{1}{ \Hcal_s} (\psi^A_s- \psi^A_o)
-\frac{2}{\Hcal_s \Delta \eta} \partial_r \psi^A_s \int_{\eta_s}^{\eta_o}\!d \eta'\psi^I\left(\eta'\right)}
\nonumber \\
& &
{ 
-\left(\frac{1}{ \Hcal_s\Delta \eta}\partial_\eta  \psi^A_s-\partial_r  \psi^A_s\right)\frac{1}{ \Hcal_s} \left[\!\psi^I_o-\psi_o^A-(\psi^I_s-
\psi_s^A)-2\int_{\eta_s}^{\eta_o}\!\!\!d\!\eta'\!\partial_{\eta'}\psi^I\left(\eta'\right)\right]
}
\nonumber \\
& &
-2 \psi^A_o \frac{1}{\Delta \eta}\int_{\eta_s}^{\eta_o} d \eta'  \psi^I\left(\eta'\right)
+\psi^A_s \frac{1}{\Delta \eta}\int_{\eta_s}^{\eta_o} d \eta'  \frac {\eta' - \eta_s}{\eta_o - \eta'}\Delta_2  \psi^I\left(\eta'\right)
{ +\frac{1}{ \Hcal_s \Delta \eta} (\psi^A_s- \psi^A_o)\frac{1}{ \Delta \eta} \int_{\eta_s}^{\eta_o}\!d \eta' \Delta_2 \psi^I\left(\eta'\right)}
 \nonumber \\
& &
+\left[\frac{1}{ \Hcal_s \Delta \eta}(-\psi_s^A+2 \psi_o^A)+\frac{1}{ \Hcal_s \Delta \eta} 
\frac{\Hcal_s'}{\Hcal_s^2}(\psi^A_s- \psi^A_o)-\psi_o^A\right]\left(-2\int_{\eta_s}^{\eta_o}\!\!\!d\!\eta'\!\partial_{\eta'}\psi^I\left(\eta'\right)\right)-\frac{1}{2} (\psi^A_o)^2+\psi^A_s \psi^A_o
 \nonumber 
 \\
& &
-(\psi^A_s)^2+\psi^A_o \psi^I_o-\psi^A_s \psi^I_o+\psi^A_o \psi^I_s-2 \psi^A_s \psi^I_s
+\frac{1}{ \Hcal_s \Delta \eta} 
\frac{\Hcal_s'}{\Hcal_s^2}\left[\frac{1}{2}(\psi^A_s- \psi^A_o)^2-\psi^A_o \psi^I_o+\psi^A_s \psi^I_o+
\psi^A_o \psi^I_s-\psi^A_s \psi^I_s\right]
\nonumber 
\\
& &
+\frac{2}{\Hcal_s  \Delta \eta}\partial_a\!  \psi^A_s
\!\int_{\eta_s}^{\eta_o}\!\!\!d\!\eta'\gamma_0^{ab} \partial_b\!\int_{\eta'}^{\eta_o}\!\!d\!\eta''\psi^I\left(\eta''\right)
+\frac{1}{\Hcal_s}\partial_a  \psi^A_s
\gamma_{0s}^{ab} \partial_b \int_{\eta_s}^{\eta_o} d \eta' \psi^I\left(\eta'\right)
\nonumber 
\\
& &
+\!\frac{4}{\Delta \eta} \int_{\eta_s}^{\eta_o} d \eta' (\psi^I\psi^A)\left(\eta'\right)
-\!\frac{2}{\Delta \eta} \int_{\eta_s}^{\eta_o}\!\!d \eta' \frac {\eta' - \eta_s}{\eta_o - \eta'}
\Delta_2  (\psi^I\psi^A)\left(\eta'\right)
\label{pathASb}
\eea

We can note how several new terms appear when we consider an anisotropic stress.
In particular, we have a new genuine second order lensing (see last term of Eq.(\ref{pathASb})), which is non zero only when we consider models of dark energy with anisotropic stress (or modified gravity models).
As a consequence,  this could be used to test these models.

\section{Comparison with previous results: vanishing anisotropic stress}
\label{Sec4}
\setcounter{equation}{0}

Let us now consider the particular case with vanishing anisotropic stress.
We begin by showing that the results for the luminosity distance-redshift relation reported in \cite{Fanizza:2013doa} agree, for this particular case, with the above results.
In the case of vanishing anisotropic stress we have $\psi^I=\psi$ and $\psi^A=0$, $\bar{d}_L$ can be
then easily obtained using Eqs. (\ref{finaldLord1})-(\ref{pathASb}), in particular we have that $\bar{\delta}_{mixed,b}^{(2)}=\bar{\delta}_{path,b}^{(2)}=0$, the form of $\bar{\delta}_{pos}^{(2)}$ does not change, and $\bar{\delta}_{mixed,a}^{(2)}$ and $\bar{\delta}_{path,a}^{(2)}$ are obtained just substituting $\psi^I$ with $\psi$.

Let us start from the explicit expressions reported in \cite{Fanizza:2013doa}:
\bea
\bar{\delta}_{pos}^{(2)} &=& \frac{1}{2} \left(1-\frac{1}{ \Hcal_s \Delta \eta}\right)\Bigg\{ \left(\partial_r P_s\right)^2 + (\gamma_0^{ab})_s \partial_a P_s \, \partial_b P_s
- \frac{2}{\Hcal_s} \left( \partial_r P_s - \partial_r P_o \right) \left( \Hcal_s \partial_r P_s + \partial_r^2 P_s \right) \nonumber \\
 &-& \int_{\eta_{in}}^{\eta_s^{(0)}} d\eta' \frac{a(\eta')}{a(\eta_s^{(0)})} \partial_r \left[ \phi^{(2)} - \psi^2 + (\partial_r P)^2 + \gamma_0^{ab} \partial_a P  \partial_b P \right] (\eta',\Delta\eta,\tilde{\theta}^a) 
\Bigg\} \nonumber \\
&+& \frac{1}{2 \Hcal_s \Delta \eta} \left\{ \left(\partial_r P_o\right)^2 +\lim_{r\rightarrow 0} \left[\gamma_0^{ab} \partial_a P \partial_b P \right]
\right. \nonumber \\
&-& \left. \int_{\eta_{in}}^{\eta_o} d\eta' \frac{a(\eta')}{a(\eta_o)} \partial_r \left[ \phi^{(2)} - \psi^2 + (\partial_r P)^2 + \gamma_0^{ab} \partial_a P \partial_b P \right] (\eta',0,\tilde{\theta}^a)\right\}
\nonumber 
\\
&-& \frac{1}{2 \Hcal_s \Delta\eta} \left( 1 - \frac{\Hcal_s'}{\Hcal_s^2} \right) \left( \partial_r P_s - \partial_r P_o \right)^2 \, ,
\label{OldForm1}
\eea
\bea
\bar{\delta}_{mixed}^{(2)} &=& \left(1-\frac{1}{ \Hcal_s \Delta \eta}\right) \Bigg\{\partial_r P_s  J_2^{(1)}-\left(
\partial_r P_s-\partial_r P_o \right) \frac{1}{\Hcal_s} \partial_\eta \psi_s - (\gamma_0^{ab})_s \partial_a Q_s \partial_b P_s \nonumber \\
&+& \frac{1}{\Hcal_s} \partial_+ Q_s \partial_r^2 P_s + Q_s  \partial_r^2 P_s  \nonumber \\
&+&
\frac{1}{2} \partial_a ( \partial_r P_s - \partial_r P_o ) \, \left( \int_{\eta_o}^{\eta_s^{(0)-}} dx ~ \left[ {\gamma}_0^{ab} ~ \partial_b Q \right] (\eta_s^{(0)+},x,\tilde{\theta}^a) \right)
\Bigg\}
\nonumber \\
&-& \frac{1}{\Hcal_s \Delta\eta}  \left(\psi_o - \psi_s - J_2^{(1)} \right) \partial_r P_o +\frac{Q_s}{\Delta \eta} \partial_r P_s
\nonumber \\
&+& \frac{1}{\Delta \eta} ( \partial_r P_s - \partial_r P_o ) \Bigg\{ \frac{1}{\Hcal_s} \left( 1 - \frac{\Hcal_s'}{\Hcal_s^2} \right) 
\partial_+ Q_s + \frac{2}{\Hcal_s} \psi_s\Bigg\} \nonumber \\
&+& \frac{1}{\Hcal_s} ( \partial_r P_s - \partial_r P_o ) \!\left\{ \partial_\eta \psi_s - \partial_r \psi_s - \frac{1}{\Delta \eta^2}\! \int_{\eta_s^{(0)}}^{\eta_o} d\eta' \Delta_2 \psi (\eta', \eta_o - \eta', \tilde{\theta}^a) \right\}\!.
\label{OldForm2}
\eea
\bea
\bar{\delta}_{path}^{(2)} &=& \left(1-\frac{1}{ \Hcal_s \Delta \eta}\right) \Bigg\{ - \frac{1}{4} \left( \phi_s^{(2)} - \phi_o^{(2)} \right) + \frac{1}{4} \left( \psi_s^{(2)} - \psi_o^{(2)} \right) + \frac{1}{2} \psi_s^2 -  \frac{1}{2} \psi_o^2 
 \nonumber \\
&-& (\psi_s + J_2^{(1)} ) \partial_+ Q_s +\frac14 (\gamma_{0}^{ab})_s \partial_a Q_s  \partial_b Q_s + Q_s \left( - \partial_+^2 Q_s + \partial_+ \psi_s \right)  
+ \frac{1}{{\mathcal H}_s} \partial_+ Q_s  
\, \partial_\eta \psi_s   \nonumber \\
&+& \frac14 \int_{\eta_o}^{\eta_s^{(0)-}} dx~ \partial_+ \left[ {\phi}^{(2)} + {\psi}^{(2)} + 4 {\psi} ~ \partial_+ Q + {\gamma}_{0}^{ab} ~ \partial_a Q ~ \partial_b Q \right] (\eta_s^{(0)+},x,\tilde{\theta}^a) \nonumber \\
&-& \frac{1}{2} \partial_a (\partial_+ Q_s) \, \left( \int_{\eta_o}^{\eta_s^{(0)-}} dx ~ \left[ {\gamma}_0^{ab} ~ \partial_b Q \right] (\eta_s^{(0)+},x,\tilde{\theta}^a) \right)
\Bigg\} \nonumber \\
&-& \frac{1}{2}\psi_s^{(2)} - \frac{1}{2} \psi_s^2 - K_2 + \psi_s J_2^{(1)} +\frac{1}{2}(J_2^{(1)})^2 
+ J_2^{(1)} \frac{Q_s}{\Delta \eta}-\frac{1}{\Hcal_s \Delta\eta} \left( 1 - \frac{\Hcal_s'}{\Hcal_s^2} \right) \frac12 (\partial_+ Q_s )^2 
\nonumber \\
&-& \frac{2}{\Hcal_s \Delta \eta} \psi_s \partial_+ Q_s + \frac12 \partial_a \left( \psi_s + J_2^{(1)} + \frac{Q_s}{\Delta \eta} \right)   \left( \int_{\eta_o}^{\eta_s^{(0)-}} dx ~ \left[ {\gamma}_0^{ab} ~ \partial_b Q \right] (\eta_s^{(0)+},x,\tilde{\theta}^a) \right) 
\nonumber
\\
&+& \frac14 \partial_a Q_s  \partial_+ \left( \int_{\eta_o}^{\eta_s^{(0)-}} dx ~ \left[ {\gamma}_0^{ab} ~ \partial_b Q \right] (\eta_s^{(0)+},x,\tilde{\theta}^a) \right)
\nonumber \\
&+& \frac{1}{16} \partial_a \left( \int_{\eta_o}^{\eta_s^{(0)-}} dx ~ \left[ {\gamma}_0^{bc} ~ \partial_c Q \right] (\eta_s^{(0)+},x,\tilde{\theta}^a) \right) \partial_b \left( \int_{\eta_o}^{\eta_s^{(0)-}} d\bar{x} ~ \left[ {\gamma}_0^{ad} ~ \partial_d Q \right] (\eta_s^{(0)+},\bar{x},\tilde{\theta}^a) \right) \nonumber \\
&-& \frac{1}{4 \Delta \eta} \int_{\eta_o}^{\eta_s^{(0)-}} dx~ \left[ {\phi}^{(2)} + {\psi}^{(2)} + 4 {\psi} ~ \partial_+ Q + 
{\gamma}_{0}^{ab} ~ \partial_a Q ~ \partial_b Q \right] (\eta_s^{(0)+},x,\tilde{\theta}^a) \nonumber \\
&+& \frac{1}{\Hcal_s} \partial_+ Q_s \left\{ - \partial_\eta \psi_s + \partial_r \psi_s + \frac{1}{\Delta \eta^2} \int_{\eta_s^{(0)}}^{\eta_o} d\eta' \Delta_2 \psi (\eta', \eta_o - \eta', \tilde{\theta}^a) \right\} \nonumber \\
&+& Q_s \left\{ \partial_r \psi_s + \partial_+ \left(\int_{\eta_o}^{\eta_s^{(0)-}} d x \frac{1}{(\eta_s^{(0)+}-x)^2} \int_{\eta_o}^x d y \Delta_2 {\psi}(\eta_s^{(0)+}, y, \tilde{\theta}^a) \right)\right. \nonumber \\
&+& \left. \frac{1}{2 \Delta\eta^2} \int_{\eta_s^{(0)}}^{\eta_o} d \eta' \Delta_2 \psi(\eta', \eta_o-\eta', \tilde{\theta}^a)
 \right\} \nonumber \\
&+& 
\frac{1}{16 \sin^2 \tilde{\theta}} \left( \int_{\eta_o}^{\eta_s^{(0)-}} dx ~ \left[{\gamma}_0^{1b} ~ \partial_b Q \right] (\eta_s^{(0)+},x,\tilde{\theta}^a) \right)^2 \, ,
\label{OldForm3}
\eea
To compare these terms with the results in Eqs. (\ref{posAS}),  (\ref{mixedASa}) and (\ref{pathASa}) (with $\psi^I=\psi$) we have to express them in a more familiar form. 
To this aim we use the results of Eqs. (\ref{PQ}) and (\ref{Us1}-\ref{Us4}) (with $\psi^I=\psi$), 
together with the following relations evaluated for the case of vanishing anisotropic stress
\be 
\partial^2_+ Q_s = \frac{3}{2}\left(\partial_\eta \psi_o-\partial_\eta \psi_s \right)+\frac{1}{2}\left(\partial_r \psi_o-\partial_r \psi_s \right) -2 \int_{\eta_s}^{\eta_o} d \eta' \partial^2_{\eta'}\psi\left(\eta'\right) 
\label{UsBi}
\ee
\bea
& & \partial_+ \left(\int_{\eta_o}^{\eta_s^{(0)-}} d x \frac{1}{(\eta_s^{(0)+}-x)^2} \int_{\eta_o}^x d y \Delta_2 {\psi}(\eta_s^{(0)+}, y, \tilde{\theta}^a) \right)=\int_{\eta_s}^{\eta_o} d \eta' \left[-\frac{1}{(\eta_o-\eta')^3} \int_{\eta'}^{\eta_o} d \eta'' \Delta_2 \psi(\eta'')
 \right. \nonumber \\ & & \left. \,\,\,\,\,\,\,\,\,\,\,\,\,\,\,\,\,\,\,\,\,\,\,\,\,\,\,\,\,\,\,\,\,\,\,\,\,\,\,\,\,\,\,\,\,\,\,\,\,\,\,\,\,\,\,\,\,\,\,\,\,\,\,\,\,+\frac{1}{(\eta_o-\eta')^2} \left(+\frac{1}{2}  \Delta_2 \psi(\eta')+
\int_{\eta'}^{\eta_o} d \eta'' \partial_{\eta''} \Delta_2 \psi(\eta'') \right)
\right]
\eea
\bea
& & \frac14 \partial_a Q_s  \partial_+ \left( \int_{\eta_o}^{\eta_s^{(0)-}} dx ~ \left[ {\gamma}_0^{ab} ~ \partial_b Q \right] (\eta_s^{(0)+},x,\tilde{\theta}^a) \right)=\partial_a \left(\int_{\eta_s}^{\eta_o} d \eta' \psi(\eta') \right)
\int_{\eta_s}^{\eta_o} d \eta' \left[\frac{2}{(\eta_o-\eta')}  \gamma_0^{ab}  \partial_b \int_{\eta'}^{\eta_o} d \eta'' \psi(\eta'')
 \right. \nonumber \\ & & \left. \,\,\,\,\,\,\,\,\,\,\,\,\,\,\,\,\,\,\,\,\,\,\,\,\,\,\,\,\,\,\,\,\,\,\,\,\,\,\,\,\,\,\,\,\,\,\,\,\,\,\,\,\,\,\,\,\,\,\,\,\,\,\,\,\,
-\gamma_0^{ab}  \partial_b \left(\psi(\eta')+2
\int_{\eta'}^{\eta_o} d \eta'' \partial_{\eta''} \psi(\eta'') \right)
\right]\,.
\eea
Furthermore $J^{(1)}_2$ is the first order lensing term and is given by
\be 
 J^{(1)}_2 
 = \frac{1}{\Delta\eta} \int_{\eta_s^{(0)}}^{\eta_o} d \eta' \,\frac {\eta' - \eta_s^{(0)}}{\eta_o - \eta'} \Delta_2 \psi(\eta', \eta_o-\eta', \tilde{\theta}^a) \,.
 \ee
This corresponds to the lowest order contribution obtained from 
 \be
J_2 = \frac{1}{2}\left[\cot \theta ~ \tilde{\theta}^{(1)}+\partial_a \tilde{\theta}^{a (1)}\right] \,,
\ee
We then have also a genuine second order lensing term given by 
\bea 
K_2 &=& \frac{1}{2}\left[\cot \theta ~ \tilde{\theta}^{(2)}+\partial_a \tilde{\theta}^{a (2)}\right] = \frac12 \nabla_a \tilde{\theta}^{a (2)} 
\nonumber \\
&=& \frac{1}{4} \frac{1}{\Delta \eta} \int_{\eta_s}^{\eta_o} d \eta' \frac {\eta' - \eta_s}{\eta_o - \eta'} \Delta_2 \left[\phi^{(2)}
\left(\eta'\right) + \psi^{(2)}\left(\eta'\right) 
+ 4 \psi\left(\eta'\right) \left(\psi_o - \psi\left(\eta'\right) -2 \int_{\eta'}^{\eta_o} d \eta'' \partial_{\eta''}\psi\left(\eta''\right)\right) \right.
\nonumber \\ & & \left.
+ 4 \gamma_0^{ab} \partial_b \left( \int_{\eta'}^{\eta_o} d \eta'' \psi\left(\eta''\right) \right) 
\partial_a \left( \int_{\eta'}^{\eta_o} d \eta'' \psi\left(\eta''\right) \right) \right]+ \int_{\eta_s}^{\eta_o} d \eta'\left\{ 
\psi\left(\eta'\right) \lim_{\bar{\eta}\rightarrow \eta_o}\left(\frac{1}{(\eta_o-\bar{\eta})^2}  \int_{\bar{\eta}}^{\eta_o} d \eta'' \Delta_2
\psi\left(\eta''\right)\right) 
\right. \nonumber \\ 
& & \left. 
-2 \psi\left(\eta'\right)\frac{1}{\eta_o-\eta'}\int_{\eta'}^{\eta_o} d \eta''  \frac {\eta'' - \eta'}{\eta_o - \eta''}\Delta_2 \partial_{\eta''} \psi\left(\eta''\right)
+2 \gamma_0^{ab} \partial_b \left(\int_{\eta'}^{\eta_o} d \eta'' \psi\left(\eta''\right)\right) \frac{1}{\eta_o-\eta'}\!\int_{\eta'}^{\eta_o}\!d \eta''  \frac {\eta''-\eta'}{\eta_o-\eta''}\partial_a \Delta_2 \psi\left(\eta''\right)
\right. \nonumber \\ 
& & \left.
-\left(\psi_o-2 \psi\left(\eta'\right)-2\int_{\eta'}^{\eta_o}\!d \eta'' \partial_{\eta''}\psi\left(\eta''\right)\right)
\frac{1}{(\eta_o-\eta')^2}\int_{\eta'}^{\eta_o}\!d \eta''\!\Delta_2 \psi\left(\eta''\right)+\partial_a \psi\left(\eta'\right)\left[
 \lim_{\bar{\eta}\rightarrow \eta_o}\left(\gamma_0^{ab} \partial_b\!\!\int_{\bar{\eta}}^{\eta_o} d \eta''
\psi\left(\eta''\right)\right) 
\right. \right. \nonumber \\ 
& & \left. \left.
-2 \int_{\eta'}^{\eta_o}\!d \eta'' \gamma_0^{ab} \partial_b \int_{\eta''}^{\eta_o}\!d \eta'''\partial_{\eta'''}\psi\left(\eta'''\right) 
\right]
+2 \partial_a \left[\gamma_0^{d b} \partial_b \int_{\eta'}^{\eta_o} d \eta'' \psi\left(\eta''\right)\right]\int_{\eta'}^{\eta_o} d \eta'' \partial_d
\left[\gamma_0^{ac} \partial_c \int_{\eta''}^{\eta_o} d \eta''' \psi\left(\eta'''\right)\right]
\right. \nonumber \\ 
& & \left.
-\gamma_0^{ab} \partial_a \left(\psi_o-2 \psi\left(\eta'\right)-2\int_{\eta'}^{\eta_o}\!d \eta'' \partial_{\eta''}\psi\left(\eta''\right)\right)
 \partial_b \int_{\eta'}^{\eta_o}\!d \eta''\psi\left(\eta''\right)
\right\} \,.
\label{UsBf}
\eea

Using Eqs. (\ref{UsBi})-(\ref{UsBf}) to evaluate Eqs. (\ref{OldForm1}), (\ref{OldForm2}) and (\ref{OldForm3}) one can prove that these are perfectly equivalent to  Eqs. (\ref{posAS}),  (\ref{mixedASa}) and (\ref{pathASa}) with $\psi^I=\psi$ and $\psi^A=0$.

In \cite{Umeh:2014ana} (see also \cite{Umeh:2012pn}) the results for the perturbed redshift and luminosity distance was also derived, but in a different way, for the particular case of vanishing anisotropic stress. In \cite{Umeh:2014ana} the authors work mainly in a perturbed Minkowski space-time and conformally transform their results  to the original FLRW space-time at the end.
Therefore, the comparison of these two independent results, the one of \cite{BenDayan:2012wi,Fanizza:2013doa} here further generalized and the one of \cite{Umeh:2014ana}, is of fundamental importance.
Considering only scalar perturbations\footnote{In \cite{BenDayan:2012wi}  also the vector and tensor perturbations were considered. The results of  \cite{BenDayan:2012wi}  are in agreement \cite{DiDioMarozzi} with the ones given in \cite{DiDio:2012bu}.}
and the different formalism used, 
we find that the results of Eqs. (\ref{Redshift1}), (\ref{Redshift2}), (\ref{Redshift1bar}) and (\ref{Redshift2bar}), for $\psi^I=\psi$ and $\psi^A=0$,
do not exactly coincide with the results of \cite{Umeh:2014ana}.
To be more precise, the first order coincides, while splitting the second order contribution of Eq. (\ref{Redshift2bar}) in a similar way as proposed in \cite{Umeh:2014ana} we find some discrepancies. Namely, we can write:
\bea
\delta^{(2)} \bar{z} &=& \delta^{(2)} z_S + \delta^{(2)} z_{SW} + \delta^{(2)} z_{SW\mbox{x}ISW} + \delta^{(2)} z_{SW\mbox{x}Dop||} + \delta^{(2)} z_{Dop||} + \delta^{(2)} z_{Dop\perp} + \delta^{(2)} z_{Dop||\mbox{x}ISW||} 
\nonumber \\
& & 
+ \delta^{(2)} z_{Dop\perp\mbox{x}ISW\perp} + \delta^{(2)} z_{IISW} +\delta^{(2)} z_{\theta^a_o}
\eea
with
\begin{eqnarray}
\delta^{(2)} z_S &=&  v^{(2)}_{||o}- v^{(2)}_{||s}+\frac{1}{2}\left(\phi_o^{(2)}-\phi_s^{(2)}\right)
-\frac{1}{2} \int_{\eta_s}^{\eta_0} d\eta' \partial_{\eta'}  \left[ \phi^{(2)}\left( \eta'\right) + \psi^{(2)}\left( \eta'\right)\right]
\\ 
\delta^{(2)} z_{SW} &=& \frac{3}{2} \psi_s^2-\frac{1}{2} \psi_o^2-\psi_s \psi_o
\\ 
\delta^{(2)} z_{SW\mbox{x}ISW} &=& 2 \left(\psi_s-\psi_o\right)  \int_{\eta_s}^{\eta_0} d\eta' \partial_{\eta'} \psi\left( \eta'\right) 
\\ 
\delta^{(2)} z_{SW\mbox{x}Dop||} &=& \left(v_{||o}- v_{||s}\right)\left(\psi_o-\psi_s\right)=\left(\psi_s v_{||s} + \psi_o  v_{||o}\right)
- \left(\psi_s v_{||o}+\psi_o v_{||s}\right)
\label{T4redshift}
\\ 
\delta^{(2)} z_{Dop||} &=& \frac{1}{2}\left(v_{||o}- v_{||s}\right)^2
\\ 
\delta^{(2)} z_{Dop\perp} &=& \frac{1}{2} \left( v^a_{\perp s} v_{\perp a\,s} 
-  v^a_{\perp o} v_{\perp a\,o}\right) 
\\ 
\delta^{(2)} z_{Dop||\mbox{x}ISW||} &=& -2\left(v_{||o}- v_{||s}\right) \int_{\eta_s}^{\eta_o} d \eta' \partial_{\eta'}\psi\left(\eta'\right)
\\ 
\delta^{(2)} z_{Dop\perp\mbox{x}ISW\perp} &=& -2 a \, v^a_{\perp s} \partial_a \int_{\eta_s}^{\eta_o} d \eta' \psi\left(\eta'\right)
\\ 
\delta^{(2)} z_{IISW} &=& 4\int_{\eta_s}^{\eta_0} d\eta' \left[\psi\left(\eta'\right) \partial_{\eta'}\psi\left(\eta'\right) 
+ \partial_{\eta'}\psi\left(\eta'\right) \int_{\eta'}^{\eta_o} d \eta'' \partial_{\eta''}\psi\left(\eta''\right)
+ \psi\left(\eta'\right) \int_{\eta'}^{\eta_o} d \eta'' \partial^2_{\eta''}\psi\left(\eta''\right)
\right.
\nonumber \\ & & \left.
- \gamma_0^{ab} \partial_a \left( \int_{\eta'}^{\eta_o} d \eta'' \psi\left(\eta''\right) \right) 
\partial_b \left( \int_{\eta'}^{\eta_o} d \eta'' \partial_{\eta''}\psi\left(\eta''\right) \right) \right] 
\label{T9redshift}
\\
\delta^{(2)} z_{\theta^a_o} &=& +2 \partial_a  \left(v_{||s}+\psi_s\right)  \int_{\eta_s}^{\eta_o}\!d \eta' \gamma_0^{ab} \partial_b \int_{\eta'}^{\eta_o}\!d \eta'' \psi\left(\eta''\right)
\nonumber \\
& & 
+4 \int_{\eta_s}^{\eta_o}\!\!d\!\eta'\partial_a \left(\partial_{\eta'}\psi\left(\eta'\right)\right)
\!\int_{\eta_s}^{\eta_o}\!\!\!d\!\eta''\gamma_0^{ab} \partial_b\!\int_{\eta''}^{\eta_o}\!d\!\eta'''\psi\left(\eta'''\right)\,.
\label{T10redshift}
\end{eqnarray}
Here $\delta^{(2)} z_S$ stands for the genuine second order terms and the other names are given in terms of the physical effects connected with the relative term (doppler effect, Sachs-Wolfe effect, integrated and double integrated Sachs-Wolfe effect).
One can then note that while the terms:  $\delta^{(2)} z_S$, $\delta^{(2)} z_{SW}$, $\delta^{(2)} z_{SW\mbox{x}ISW}$, $\delta^{(2)} z_{Dop||}$,  $\delta^{(2)} z_{Dop\perp}$, $\delta^{(2)} z_{Dop||\mbox{x}ISW||}$, and $\delta^{(2)} z_{Dop\perp\mbox{x}ISW\perp}$  coincide with the corresponding terms of \cite{Umeh:2014ana}, 
the terms: $\delta^{(2)} z_{SW\mbox{x}Dop||}$ and $\delta^{(2)} z_{IISW}$ seem to differ from the corresponding terms of \cite{Umeh:2014ana}. 
Just to make an example, in the term $\delta^{(2)} z_{IISW}$  a double partial derivative with respect to the  conformal time is present, while this is not present in the corrispondent term of \cite{Umeh:2014ana}.

The term $\delta^{(2)} z_{\theta^a_o}$ comes from the fact that 
we want to write the perturbed redshift as a function of the observer's angular coordinates (see Sec. \ref{Sec3A}), which differ from the angular coordinates at the source ($\theta^a_s\neq \theta^a_o$ for $\eta_s\neq \eta_o$, see Eq.(\ref{thetatilde1orderShort_Inverted})).
This term appears to be absent in \cite{Umeh:2014ana}, suggesting that in \cite{Umeh:2014ana} the observed redshift is not written as a function of the direction of observation.


Moving to the luminosity distance-redshift relation,
the comparison of the results in Eqs. (\ref{posAS}),  (\ref{mixedASa}) and (\ref{pathASa})  (with $\psi^I=\psi$ and $\psi^A=0$) with the ones of \cite{Umeh:2014ana} is very involved, in fact results that are equal can look different by using a simple integration by parts. On the other hand, we can try to compare terms with the same physical meaning and which can be isolated easily from the other terms. 
As an example, considering only the term $\sim v_{||s}^2$, in $\bar{\delta}_{pos}^{(2)}$, we have 
\be
\left[-\frac{1}{2}+\frac{1}{2}\frac{1}{\Hcal_s \Delta \eta}\frac{\Hcal_s'}{\Hcal^2_s}\right] v_{||s}^2 \,, 
\label{Termv2us}
\ee
while if we look at Eq.(140) of  \cite{Umeh:2014ana} we have
\be
\left[+\frac{1}{2}-\frac{1}{\Hcal_s \Delta \eta}+\frac{1}{2}\frac{\Hcal_s'}{\Hcal^2_s}\right] v_{||s}^2\,. 
\label{Termv2UmehetAl}
\ee
Clearly the quantities appearing in Eqs. (\ref{Termv2us}) and (\ref{Termv2UmehetAl}) are different.
Therefore, from this simple check, and from the fact that also the second order perturbed redshift of \cite{Umeh:2014ana} differs from our expression of Eq.(\ref{Redshift2bar}), we conclude that the recent results in \cite{Umeh:2014ana} do not agree with the results of \cite{BenDayan:2012wi,Fanizza:2013doa}  and with the ones here presented.

\section{Conclusions}
\label{Sec5}
\setcounter{equation}{0}

Let us summarize the results of this work and make some further comment.

The main results of the paper are the perturbed redshift and luminosity distance-redshift relation up to second order in perturbation theory in Poisson gauge, obtained by generalizing the results in \cite{BenDayan:2012wi,Fanizza:2013doa} to include an
anisotropic stress. 
These results, presented in Eqs. (\ref{Redshift1}), (\ref{Redshift2}), (\ref{Redshift1bar}) and (\ref{Redshift2bar}), and
in Eqs. (\ref{finaldLord1})-(\ref{pathASb}), are therefore valid for general dark energy models and (most) modified gravity models, for which a non-vanishing anisotropic stress frequently appears.
The results are presented using a standard formalism close to the one introduced in \cite{Bonvin:2005ps}. 

The evaluation of LSS observables in models with anisotropic stress is of fundamental importance for the understanding of the dark 
energy problem. In fact, a direct detection of an observational signature of anisotropic stress will rule out $\Lambda$CDM. 
Taking into consideration the anisotropic stress several new terms appear in Eqs. (\ref{finaldLord1})-(\ref{pathASb}). 
In particular,  in Eq.(\ref{pathASb}) a new genuine second order lensing term (last one in (\ref{pathASb})) appears.
This is non zero only for models of dark energy with anisotropic stress (or for modified gravity models), and,
as a consequence,  could be used to test alternative models of dark energy.

Furthermore, we have seen that the results obtained in Eqs. (\ref{finaldLord1})-(\ref{pathASb}) reduce to the ones given in 
\cite{BenDayan:2012wi,Fanizza:2013doa} for the particular case of vanishing anisotropic stress.

To this aim we use a series of results (Eqs. (\ref{PQ}), (\ref{Us1}-\ref{Us4}) and (\ref{UsBi}-\ref{UsBf})) which constitute a dictionary to 
translate the quantities that one usually obtains when going from the GLC gauge to the Poisson gauge, to more standard variables.

We have then summarized, and in part illustrated, the innovative approach used in \cite{BenDayan:2012wi,Fanizza:2013doa} to write LSS observables (like redshift and luminosity distance) to a given order in perturbation theory starting from the GLC gauge. This innovative approach enormously simplifies
this task and can be used also to obtain other useful observable like the galaxy number counts at second order \cite{DDMM} 
(see \cite{YFZandY,NC} for the first order case, and \cite{NCothersOrd2} for others recent results at second order).

Finally, we have partially compared  our results about the perturbed redshift and luminosity-redshift relation, with those of \cite{Umeh:2014ana} (see also \cite{Umeh:2012pn}) for the case of vanishing anisotropic stress, showing that there is still some disagreement with the results presented here and in \cite{BenDayan:2012wi,Fanizza:2013doa}.
We stress that arriving at a commonly accepted expression for the perturbed redshift and luminosity-redshift relation is of fundamental importance in view of the future cosmological surveys. 
In fact, any uncertainty in the theoretical description of the observed redshift and luminosity distance will impact the reliability
of the interpretation of our present and future cosmological observations.
We do hope that this work will stimulate further investigation in this direction.

\section*{Acknowledgments}

We wish to thank Ruth Durrer, Maurizio Gasperini and Gabriele Veneziano for reading a preliminar version of this manuscript and for the many and fruitful comments and suggestions, and Enea Di Dio for useful discussions.

GM was partially supported by the Marie Curie IEF, Project NeBRiC - ``Non-linear effects and backreaction in classical and quantum cosmology".


\end{document}